\documentclass[reprint,twocolumn,superscriptaddress,showpacs,nofootinbib,notitlepage,preprintnumbers,aps,prd,floatfix]{revtex4-2}
\usepackage[utf8]{inputenc}
\usepackage{graphicx}
\usepackage{latexsym,amsmath,amssymb,lmodern,float,url}
\usepackage{qcircuit}
\usepackage{natbib}
\usepackage{color}
\usepackage{microtype}
\usepackage{bbold}
\usepackage{slashed}
\usepackage{multirow}
\usepackage{tikz}
\usepackage{xcolor}
\usepackage{mathrsfs}  
\usetikzlibrary{shapes}
\usepackage{adjustbox}
\usepackage{todonotes}
\usepackage{braket}
\usepackage{nicematrix}
\usepackage{mathtools}

\usepackage[colorlinks=true,backref=false, linktocpage=true,
citecolor=blue,urlcolor=blue,linkcolor=blue,pdfpagemode=UseOutlines]{hyperref}
\usepackage{cleveref}
\Crefname{equation}{Eq.}{Eqs.}
\hypersetup{%
 bookmarksnumbered=true,
 pdftitle = {},
 pdfsubject = {},
 pdfauthor = {},
 pdfkeywords = {}
}

\let\Re\undefined
\let\Im\undefined

\DeclareMathOperator{\Re}{Re}
\DeclareMathOperator{\Im}{Im}

\newcommand{\bfi}{\mathbf{i}}
\newcommand{\bfj}{\mathbf{j}}
\newcommand{\bfk}{\mathbf{k}}
\newcommand{\bfl}{\mathbf{l}}

\def \Im {\mathop{\rm Im}\nolimits}

\definecolor{vermillion}{RGB}{213,94,0}
\definecolor{cblue}{RGB}{0,114,178}
\definecolor{corange}{RGB}{230,159,0}
\definecolor{blgreen}{RGB}{0,158,115}
\definecolor{repurple}{RGB}{204,121,167}

\definecolor{mycolor}{RGB}{0,255,0}

\begin{document}
 \preprint{FERMILAB-PUB-25-0143-SQMS-STUDENT-T}
 
\title{The 2T-quoctit: a two-mode
bosonic qudit for high energy physics}
\author{Do\~ga Murat K\"urk\c{c}\"uo\~glu}
\email{dogak@fnal.gov}
\affiliation{Superconducting and Quantum Materials Systems Center (SQMS), Batavia, Illinois, 60510, USA.}
\affiliation{Fermi National Accelerator Laboratory, Batavia, Illinois, 60510, USA}

\author{Henry Lamm}
\email{hlamm@fnal.gov}
\affiliation{Superconducting and Quantum Materials Systems Center (SQMS), Batavia, Illinois, 60510, USA.}
\affiliation{Fermi National Accelerator Laboratory, Batavia, Illinois, 60510, USA}

\author{Oluwadara Ogunkoya}
\email{ogunkoya@fnal.gov}
\affiliation{Superconducting and Quantum Materials Systems Center (SQMS), Batavia, Illinois, 60510, USA.}
\affiliation{Fermi National Accelerator Laboratory, Batavia, Illinois, 60510, USA}

\author{Leonardo Pierattelli}
\email{leonardo.pierattelli@sns.it}
\affiliation{NEST, Istituto Nanoscienze-CNR and Scuola Normale Superiore, Piazza San Silvestro 12, I-56127 Pisa, Italy}

\date{\today}
\begin{abstract}
In this work, we study a two-mode bosonic encoding of a quoctit inside a  non-Abelian group-structured constellation of coherent states. This work is motivated by the importance of nonabelian symmetry in particle physics and the desire to have transversal nonabelian logical gates.  We use the previously developed $2T$-constellation of states used to encode a so-called $2T$-qutrit. The fidelity of the $2T$-quoctit is benchmarked against other bosonic qudits for different noise models and find it compare favorably when power constraints are considered. This paves the way for the construction of higher-dimensional qudits (e.g. a quicosotetrit with $2T$ group structure) in bosonic systems with practical applications in quantum simulations of particle physics.
\end{abstract}
\maketitle
\section{Introduction}

The 50 year history of solving lattice gauge theories (LGTs) on classical computers with Monte Carlo methods in Euclidean time has many successes. Alas, this approach struggles with real-time dynamics or with matter at finite density due to sign problems~\cite{Aarts_2016,Gattringer:2016kco,Alexandru:2020wrj,Troyer:2004ge}. While quantum computers may investigate these problems~\cite{Klco:2021lap,Banuls:2019bmf,Bauer:2022hpo,DiMeglio:2023nsa}, their finite nature preclude quantum computers from simulating the dynamics of fields defined in the infinite-volume spacetime continuum.  Therefore LGT approaches will remain relevant. With this, the fields have to be \emph{digitized} i.e. restricted to a finite set in some basis; this leads to new renormalization questions and therefore efficient digitization to the limited quantum memory is crucial.

One digitization is the \textit{discrete subgroup approximation}~\cite{Bender:2018rdp,Hackett:2018cel,Alexandru:2019nsa,Yamamoto:2020eqi,Ji:2020kjk,Haase:2020kaj,Carena:2021ltu,Armon:2021uqr,Gonzalez-Cuadra:2022hxt,Charles:2023zbl,Irmejs:2022gwv,Gustafson:2020yfe, Bender:2018rdp,Hackett:2018cel,Alexandru:2019nsa,Yamamoto:2020eqi,Ji:2020kjk,Haase:2020kaj,Carena:2021ltu,Armon:2021uqr,Gonzalez-Cuadra:2022hxt,Charles:2023zbl,Irmejs:2022gwv, Hartung:2022hoz,Carena:2024dzu,Zohar:2014qma,Zohar:2016iic,Davoudi:2024wyv} which replaces the continuous gauge group by a crystal-like discrete subgroup. This method was explored in the 70's and 80's to reduce memory and runtime on classical computers~\cite{Creutz:1979zg,Weingarten:1980hx,Weingarten:1981jy,Creutz:1982dn,Bhanot:1981xp,Petcher:1980cq,Bhanot:1981pj} and recent studies have been performed in connection with understanding the systematic approximation for quantum simulations~\cite{Hackett:2018cel,Alexandru:2019nsa,Ji:2020kjk,Ji:2022qvr,Alexandru:2021jpm,Carena:2022hpz,Calliari:2025igs}. While $U(1)$ can be replaced by any $\mathbb{Z}_n$, 
for non-Abelian groups such subgroups are finitely many: for instance, $SU(2)$ has three crystal-like subgroups, the 24-element \textit{binary tetrahedral group} $2T$, the 48-element \textit{binary octahedral group} $2O$, and the 120-element \textit{binary icosahedral group} $2I$.\footnote{These groups are sometimes denoted $\mathbb{BT}, \mathbb{BO}, \mathbb{BI}$ respectively.} Additionally, of interest here is the 8-element quaternions, $Q_8$, another subgroup of $SU(2)$. The discrete subgroup approximation naturally maps the group elements to a finite set of integers while preserving a remnant gauge symmetry which reduces the need for real number arithmetic and can integrate with quantum error correction~\cite{rajput2021quantum,Carena:2024dzu}.

With any digitization, minimizing the number of quantum registers (i.e. qubits or qudits) the field variables are mapped to is desirable in order to reduce connectivity and enhance noise robustness.~\cite{Gustafson:2023swx}. In the case of discrete groups, the best case would be to encode each gauge link into a single $d$-dimensional qudit register where $d$ is the size of the group.  Due to the mapping from nonabelian groups to the integers being nonlinear, mappings that preserve that structure could reduce the gate complexity when states stored in registers are transformed under the group theoretic operations~\cite{Alam:2021uuq,Gustafson:2022xdt,Gustafson:2023kvd,Lamm:2024jnl,Gustafson:2024kym,Murairi:2024xpc}.

Both these problems can be addressed by using qudits constructed from the bosonic modes like those proposed for superconducting radio frequency (SRF) cavities~\cite{Chakram:2021bxb,Eickbusch:2021uod,2067315,Reineri:2023ebw,Stein:2023yml,You:2024oru,Roy:2024uro,Kim:2025ywx}. First, the coherence time of these cavities can exceed 1~$ms$~\cite{Milul:2023nbc,Kim:2025ywx} allowing for potentially very large dimensional qudits. Second, suitable \textit{bosonic codes} could store such information robustly, for example by encoding the \textit{logical qudits} inside the much larger Hilbert spaces of the bosonic mode. Last but not least, using bosonic modes enables one to consider different finite encodings according to one's needs, for example by choosing Hilbert spaces with specific \textit{group structures} through concatenation of multiple modes.

This last fact, that natural encodings exist for specific group structures suggests exploration into their use for LGTs. In particular, if we can implement into a code the structure of the crystallike groups $H$ that approximate relevant gauge theory, we would know how to transform between the different group elements $h$ almost ``for free", without the need for engineering suitable sequences of quantum gates in order to perform them. To do this, we associate each $h$  a logical state embedded in the set of states of a larger group $G\supset H$ which is physically encoded in hardware.

For instance, any element of $SU(2)$ (and thus its discrete subgroups) can be defined by two complex numbers, and thus mapping these constellations to states defined in the 2d phase space of a single bosonic mode would require partitioning the space in an unnatural way.  Instead, there is a natural embedding of $SU(2)$ into two bosonic modes which allows for the $4$ real coefficients.  Then, the $4$ coefficients are associated to the 2 complex coefficients needed to describe the $2$ coherence parameters $\alpha,\beta$ of a state of the bosonic field of the type $\ket{\alpha}\!\ket{\beta}$, i.e. a $2$-mode coherent state. In general, these methods implies that different nonabelian groups are naturally represented by \textit{multimode} bosonic codes.

Specific groups can be easier to implement, mostly depending upon whether the desired group is \textit{Abelian}, like $\mathbb{Z}_n$, or \textit{non-Abelian}, like the $2T$ group itself.
For instance, several encodings exist for codewords having $\mathbb{Z}_n$ group structure, but implementing codes with non-Abelian group structure is less straightforward.
Therefore, before delving into the task of finding a multimode bosonic code that is suitable for representing crystal-like subgroups of $SU(N)$ (which are all non-Abelian), here we wish to explore the complications which arise from requiring non-Abelian \textit{group codes} and whether they present lower or higher error tolerances with respect to Abelian group codes. In this work, we revisit using $2T=\mathbb{Z}_3\rtimes Q_8$ as the larger group which was already used to construct an abelian qutrit~\cite{Denys:2022iyj}.  Here we explore how to implement a nonabelian logical quoctit ($d=8$ qudit) based on $Q_8$. 

This work is organized as follows. We briefly review the basics of bosonic error correcting codes in Sec.~\ref{sec:bosoniccodes}.  This leads into Sec.~\ref{sec:2tqudit} were we define characteristics of the $2T$-qudits.  Numerical studies of the robustness of the $2T$-quoctit are presented in Sec.~\ref{sec:numresult} and compared to the $2T$-qutrit and other bosonic codes. We conclude with discussion and future work in Sec.~\ref{sec:con}.

\section{Basics of Bosonic Codes}
\label{sec:bosoniccodes}
\noindent
Encoding quantum information in Hilbert space containing at least one bosonic mode, i.e. an oscillator, like one mode of the electromagnetic (EM) field inside a cavity, requires the use of a bosonic code.
A \textit{code} is a map between between the logical states we want and distinct subsets of possible configurations of the system.
A practical way to implement bosonic codes is by considering a finite dimensional subspace -- called the code constellation -- within the Hilbert space of \textit{coherent states} of the oscillators.
Coherent states are defined as the eigenstates of the annihilation operator of the corresponding oscillator:
\begin{equation}
\label{eq:cohstate}
    \hat{a}\ket{\alpha} = \alpha\ket{\alpha}.
\end{equation}
We call the eigenvalue $\alpha \in \mathbb{C}$, the \textit{coherence parameter} of the state.
This parameter determines the coordinates of the state in the \textit{phase space} of the oscillator's canonical variables via the displacement operator $\hat{D}(\alpha)$:
\begin{equation}
    \ket{\alpha} = 
    e^{-\frac{|\alpha|^2}{2}}
        \sum_{n=0}^{+\infty}
        \frac{\alpha^n}{\sqrt{n!}}\ket{n} =
    \hat{D}(\alpha)\ket{0}\label{eq:coherent_state}
\end{equation}
where $\hat{D}(\alpha)=e^{\alpha \hat{a}^{\dagger}-\alpha^*\hat{a}}$, and  $\{\ket{n}, n=0,1,\ldots\}$ are the \textit{Fock state} basis of the oscillator's Hilbert space. 
One important fact about coherent states is that the inner product between them is not orthogonal:
\begin{equation}
\braket{\alpha|\beta}
=e^{-\frac{|\alpha|^2}{2}-\frac{|\beta|^2}{2}+\alpha^* \beta}.
\label{eq:cohscalprod}
\end{equation}

\noindent To design a bosonic code, one first chooses a set of coherent states called the \emph{code constellation}. One simple choice is the so-called \textit{cat code}~\cite{Cochrane:1998ft}, which considers a constellation of $N$ coherent states placed on a circle of radius $|\alpha|$ in the phase space of one mode corresponding to $G=\mathbb{Z}_N$. Another notable example of bosonic codes is the \textit{Gottesman-Kitaev-Preskill} (GKP) code~\cite{Gottesman:2000di}, where the constellation consists of states encoded in a rectangular grid in phase space. 

\begin{figure}
    \centering
                \includegraphics[width=0.45\linewidth]{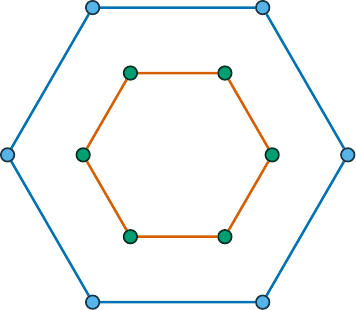}\hspace{0.25cm}
                        \includegraphics[width=0.45\linewidth]{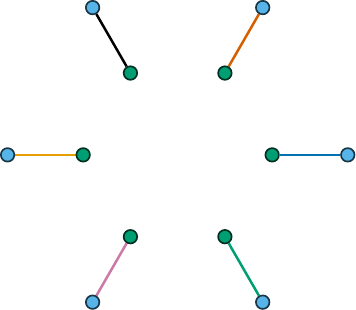}
    \caption{Example of two bosonic codes built from $\mathbb{Z}_6\times\mathbb{Z}_6$ constellation in two modes. The blue and green points denote the single-mode states $\ket{\alpha_i}\ket{0}$ and $\ket{0}\ket{\alpha_j}$ respectively. Connected points correspond to (left) the 2 states of a qubit and (right) the 6 states of a quhexit.}
    \label{fig:constellation_exp}
\end{figure}

\vspace{2mm}
\noindent To establish a correspondence between the logical states and the system's configurations, one subdivides the code constellation. From each one of these subspaces, one specific superposition is chosen as a \emph{codeword}, which represents one logical state (see Fig.~\ref{fig:constellation_exp}). The leftover Hilbert subspace would then be used for redundancy and for detecting error syndromes.  In the cat code, the $\mathbb{Z}_N$ element are divided in $\mathbb{Z}_M$ codewords; this leads to a logical $d=M/N$ dimensional qudit.

These single-mode cat qudits can be generalized to $d$ logical states of the form:
\begin{equation}
    \ket{\bar{k}}=\sum_{\ell =0}^{d\cdot n-1}
    e^{-\frac{2\pi i}{dn}(\ell \cdot k)}
    \ket{\alpha e^{\frac{2\pi i}{dn}(\ell \cdot k)}}.
\end{equation}
with the constellation being that of a \textit{Phase-Shift Keying modulation}. Following~\cite{Denys:2022iyj}, we call these the \textit{$dn$-PSK} encodings for a $d$-dimensional logical state storied in $dn$ constellation, and  we  will use them as a   benchmark in the following section.

An obvious extension is to take constellations of sets of coherent states in  more than one bosonic mode, thus defining \textit{multimode bosonic codes}.
In this case, the larger space of coherent states allows for more structurally complicated codes. One class of such codes are \textit{quantum spherical codes} where a one-to-one correspondence between every state in the constellation and one vertex of a complex polytope inscribed inside a multidimensional sphere can be made~\cite{Jain:2023deu}.
This way, the symmetries of the complex polytope associated to the chosen spherical code could be easily expressed as unitary transformations acting on the modes.  Another interesting class, \emph{group-covariant codes} are motivated by desiring transversal gate sets with special group structure~\cite{Denys:2023syu,Kubischta:2023nlb,Denys:2023syu,Lin:2024utr,Kubischta:2024fuf}.  Clearly if such codes are possible, then HEP -- where such group structure is critical -- could benefit, and we investigate here a specific bosonic code yielding a simple nonabelian qudit. 

\section{\texorpdfstring{$2T$}{2T}-Qudits}
\label{sec:2tqudit}
\noindent As a step toward larger groups, we focus on obtaining qudits from a $2T$ constellation of $24$ elements. Physically, these elements will be states in the phase space of two bosonic modes. Using this constellation, the encoding of a \textit{qutrit} was previously developed~\cite{Denys:2022iyj}. Here we investigate a \textit{quoctit} possessing ``complementary" characteristics.

\begin{figure}
    \centering
        \includegraphics[width=\linewidth]{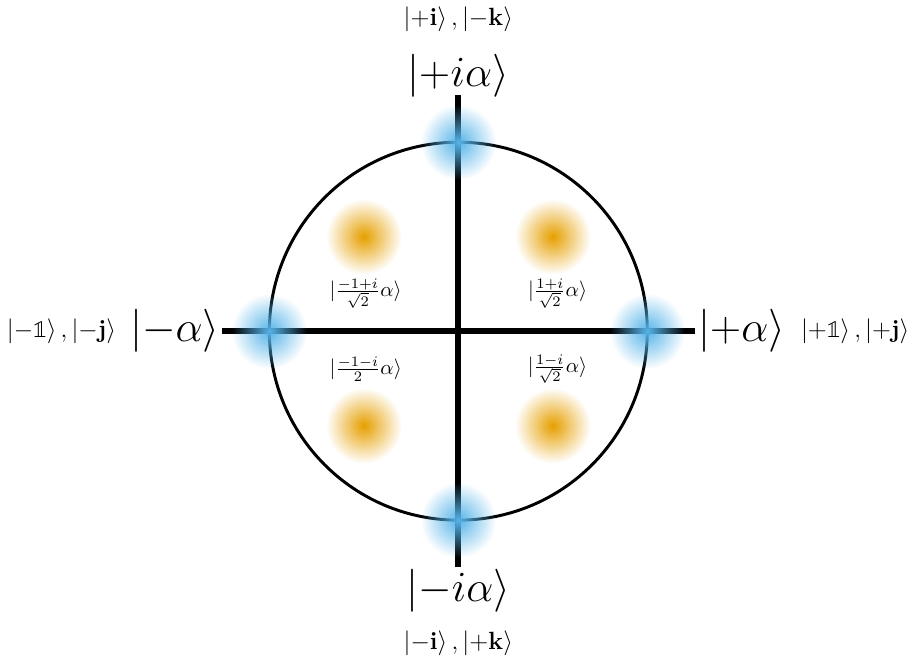}
    \caption{The $2T$ constellation states. 4 single-mode states are supported at a radius of $|\alpha|$, while 16 states of the form $\ket{\alpha}\!\ket{\beta}$ are located around $|\alpha|/\sqrt{2}$. The larger axis labels indicate coherent states, while the smaller ones denote the $2T$ element stored in each mode.}
    \label{fig:2tconstellation}
\end{figure}

The $2T$ group is a finite subgroup of the unit hypersphere in the quaternion ring $\mathbb{H}$:
\begin{equation}
    \left\{
        \pm \mathbb{1},\, \pm \bfi,\, \pm \bfj,\, \pm \bfk,\,
        \frac{1}{2}
        \left(
            \pm \mathbb{1} \pm \bfi \pm \bfj \pm \bfk
        \right)
    \right\},
    \label{eq:2tquat}
\end{equation}
with all the possible sign combinations.
Alternatively, every element can be written as the following ordered product of four generators following~\cite{Gustafson:2022xdt}:
\begin{equation}
    \left(-1\right)^m
    \bfi^n \bfj^p \bfl^q,
\end{equation}
with $m,n,p\in\{0,1\};\, q\in\{0,1,2\}$, and 
\begin{equation}
    \bfl = -\frac{1}{2}
        \left(
            \mathbb{1}+\bfi+\bfj+\bfk
        \right),
\end{equation}
Additional useful relations are:
\begin{equation}
\label{eq:genrel}
\begin{split}
&\mathbf{i}^2=\mathbf{j}^2=-\mathbb{1},~ \mathbf{l}^3=\mathbb{1}\\
&\mathbf{i} \mathbf{j} = \mathbf{k}, ~ \mathbf{j} \mathbf{k} = \mathbf{i}, ~ \mathbf{k} \mathbf{i} = \mathbf{j},~\mathbf{l}\mathbf{i}=\mathbf{j}\mathbf{l},~\mathbf{l}\mathbf{j}=\mathbf{k}\mathbf{l},~\mathbf{l}\mathbf{k}=\mathbf{i}\mathbf{l}.\\
\end{split}
\end{equation}
The correspondence between the elements in Eq.~(\ref{eq:2tquat}) and the 2-mode coherent states $\ket{\alpha}\ket{\beta}$ in the constellation is:
\begin{equation}
    a+b\bfi+c\bfj+d\bfk  \mapsto 
    \ket{(a+bi)\alpha}\!\ket{(c-di)\alpha},
\end{equation}
The quaternion coefficients will be only those of Eq.~(\ref{eq:2tquat}), and $\alpha$ is a complex factor which sets the overall coherence parameter. Given these definitions, one can define the \textit{$2T$ constellation} with a Hilbert space $\mathcal{H}_{2T}$:
\begin{align}
    \mathcal{H}_{2T}=
    \operatorname{span}&
    \bigg\{
        \ket{e^{ik\pi/2}\alpha}\!\ket{0}\!,\;
        \ket{0}\!\ket{e^{il\pi/2}\alpha}\!,\;\notag\\&\hspace{-1.2em}
        \ket{\tfrac{e^{i(2k+1)\pi/4}}{\sqrt{2}}\alpha}\!
            \ket{\tfrac{e^{i(2l+1)\pi/4}}{\sqrt{2}}\alpha}\!:
        0\leq k,l\leq 3
    \bigg\},
\end{align}
whose (non-orthogonal) basis states are shown in Fig.~\ref{fig:2tconstellation}.

A qudit can be defined by taking a $d$-independent subspaces of $\mathcal{H}_{2T}$, pick an element from each of the independent subspaces, and use them as logical states. Natural choices for these subspaces is to chose subgroups of $2T$. In fact, a $d$-dimensional qudit can be constructed out of any group $G$ whenever a subgroup $H$ exists such that for some $g\in G$ there are cosets of the form $g^k H$ for $k=0, \ldots, d-1$, and $g^d H=H$. Since $2T=\mathbb{Z}_3\rtimes Q_8$, we have two natural options for $H$:
\begin{equation}
    \mathbb{Z}_3=\{1,\bfl,\bfl^2\},\quad Q_8=\{\pm \mathbb{1},\pm \bfi,\pm \bfj,\pm \bfk\},
\end{equation} 
We first review the procedure for $\mathbb{Z}_3$, then extend to $Q_8$. 

\subsection{\texorpdfstring{$2T$}{2T}-Qutrit}

\noindent The $\mathbb{Z}_3$ subgroup naturally lends itself to defining the $2T$-qutrit via the association of $H=Q_8$ and $g=\bfl$ (the generator of $\mathbb{Z}_3$).
Using the shorthand notation $\ket{q}$ for states representing the quaternion $q=a+b\bfi+c\bfj+d\bfk$, we define (up to a normalization factor $\nu$):
\begin{equation}
    \ket{\phi_{0}}=\nu\sum_{q\in Q_8}\ket{q},\;\;
    \ket{\phi_{1}}=\nu\sum_{q\in Q_8}\ket{\bfl q},\;\;
    \ket{\phi_{2}}=\nu\sum_{q\in Q_8}\ket{\bfl^2q}.
    \label{eq:qtritwords}
\end{equation}

\begin{figure}[ht]
    \centering
    \includegraphics[width=0.48\linewidth]{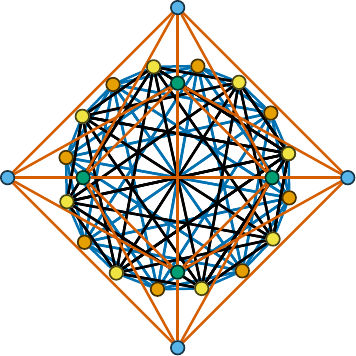}\hspace{0.25cm}
        \includegraphics[width=0.48\linewidth]{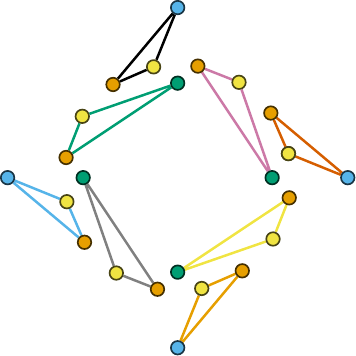}
    \caption{Representation of different multimode bosonic codes built from the 2T constellation. The outermost blue and innermost green points denote single-mode states, while the middle-radius points represent entangled two-mode states.  Connected points correspond (left) to the 3 states of the $2T$-qutrit and (right) to the 8 states of the $2T$-quoctit.}
    \label{fig:multicode}
\end{figure}

In other words, the $2T$ constellation is divided into three codeword constellations, each one constituted by states which represent different cosets of $2T$ of order $8$:
\begin{equation}
    \begin{aligned}
        Q_8 = &
        \left\{
            \pm \mathbb{1},\pm \bfi,\pm \bfj,\pm \bfk
        \right\}, \\
         \bfl Q_8 = &
        \left\{
            \pm\bfl,
            \pm\bfl\bfi,
            \pm\bfl\bfj,
            \pm\bfl\bfk
        \right\}, \\
         \bfl^2Q_8 = &
        \left\{
            \pm\bfl^2,
            \pm\bfl^2\bfi,
            \pm\bfl^2\bfj,
            \pm\bfl^2\bfk
        \right\}, \\
    \end{aligned}
\end{equation}
while the three codewords differ by the power of $\bfl$.
In this sense, each codeword has the $Q_8$ group structure, while the \textit{code} has the \textit{Abelian} group structure of $\mathbb{Z}_3$.
Therefore, the three $\ket{\phi_{n}}$ states could constitute a \textit{quantum spherical code} by themselves, but to build a qutrit, we need the orthonormal logical states.  A schematic view of this can be seen in the left of Fig.~\ref{fig:multicode}.
The previous prescription constructs an orthonormal basis of logical states $\ket{\Bar{k}}$:
\begin{equation}\label{eq:logqtrit}
    \ket{\Bar{k}} = \nu_k\sum_{n=0}^2
    e^{-\frac{2\pi i}{3}(k\cdot n)}\ket{\phi_{n}}.
\end{equation}
where
\begin{align}
    \nu_0=&\frac{1}{3(1+2\braket{\phi_1|\phi_2})};\quad
    \nu_1=\nu_2=\frac{1}{3(1-\braket{\phi_1|\phi_2})}
\end{align}
In this code, a logical $\Bar{Z}\ket{\Bar{k}}=e^{-\frac{2\pi i}{3}k}\ket{\Bar{k}}$ exists~\cite{Denys:2022iyj}. It was identified with a Gaussian passive transformation between the two modes:
\begin{equation}
    \mathcal{U}=\exp{
        \frac{2\pi i}{3\sqrt{3}}
        [
            -\hat{a}^\dagger\hat{a}
            +(1-i)\hat{a}^\dagger\hat{b}
            +(1+i)\hat{a}\hat{b}^\dagger
            +\hat{b}^\dagger\hat{b}
        ]
    }.
\end{equation}
For the two modes, we see can write the effect of $\mathcal{U}$ on coherent states $\ket{\alpha}\!\ket{\beta}$ and $\ket{\alpha'}\!\ket{\beta'}$  is $\Bar{Z}\begin{bsmallmatrix}\alpha\\\beta\end{bsmallmatrix}=\begin{bsmallmatrix}\alpha'\\\beta'\end{bsmallmatrix}$. In this basis, we find
\begin{equation}
    \mathcal{U}=-\frac{1}{2}
    \begin{bmatrix}
        1+i &   -1-i    \\
        1-i &   1-i     \\
    \end{bmatrix},
\end{equation}
which is a matrix representation of the quaternion $\bfl$ as can be seen by the following mapping:
\begin{equation}
    q=a+b\bfi+c\bfj+d\bfk\in\mathbb{H}\mapsto
    \begin{bmatrix}
        a+bi    &   -c-di   \\
        c-di    &   a-bi
    \end{bmatrix} \in \mathcal{M}_2(\mathbb{C}).
\end{equation}
Thus $\mathcal{U}$ can be understood to implement
\begin{equation}
\label{eq:uzgo}
   \mathcal{U}\ket{q}=\ket{\bfl q} \implies\mathcal{U}\ket{\phi_{n}}=\ket{\phi_{{n+1}}},
\end{equation}
i.e. it \textit{cycles} between the different codewords, just like the quaternion (left) multiplication by $\bfl$ would do. Using Eq.~(\ref{eq:logqtrit}), we directly see  $\mathcal{U}\ket{\Bar{k}}=e^{-\frac{2\pi i}{3}k}\ket{\Bar{k}}$ which is the definition of the qutrit $\bar{Z}$.

\subsection{\texorpdfstring{$2T$}{2T}-Quoctit}
\noindent A great help in constructing the qutrit came from the fact that the codeword constellations were related to each other by multiplication by the generator $\bfl$ of the $\mathbb{Z}_3$ group.  This ensured orthogonality of the code words. No such quoctit construction occurs using the $2T$ constellation, since $Q_8$ is non-cyclic and non-Abelian. Thus, we will instead start with a natural-seeming set of eight codewords (up to normalization factors):
\begin{equation}
\label{eq:initdef}
    \ket{\phi_q}=\ket{q}+\ket{q\bfl}+\ket{q\bfl^2},
    \quad\forall q\in Q_8.
\end{equation}

We construct an ortho\textit{normal} basis inside the space $\mathcal{H}_Q=\operatorname{span}\left\{ \ket{\phi_q} \right\}_{q\in Q}$ by computing the Gram matrix $\eta_{qq'}=\braket{\phi_q|\phi_{q'}}$, using the inner product defined in Eq.~(\ref{eq:cohscalprod}):
\begin{equation}
    \eta_{qq'}=
    \begin{pmatrix}
        \rho    &   \tau    &   \tau    &   \tau    &   \chi    &   \tau^*  &   \tau^*  &   \tau^*  \\
        \tau^*  &   \rho    &   \tau    &   \tau^*  &   \tau    &   \chi    &   \tau^*  &   \tau    \\
        \tau^*  &   \tau^*  &   \rho    &   \tau    &   \tau    &   \tau    &   \chi    &   \tau^*  \\
        \tau^*  &   \tau    &   \tau^*  &   \rho    &   \tau    &   \tau^*  &   \tau    &   \chi    \\
        \chi    &   \tau^*  &   \tau^*  &   \tau^*  &   \rho    &   \tau    &   \tau    &   \tau    \\
        \tau    &   \chi    &   \tau^*  &   \tau    &   \tau^*  &   \rho    &   \tau    &   \tau^*  \\
        \tau    &   \tau    &   \chi    &   \tau^*  &   \tau^*  &   \tau^*  &   \rho    &   \tau    \\
        \tau    &   \tau^*  &   \tau    &   \chi    &   \tau^*  &   \tau    &   \tau^*  &   \rho
    \end{pmatrix},
\end{equation}
where $\rho,\tau,\chi$ are functions of $\alpha$:
\begin{align}
    \rho=&3+3e^{-\frac{3+i}{2}|\alpha|^2}\left(1+e^{i|\alpha|^2}\right)\notag\\
    \tau=&e^{-\frac{3+i}{2}|\alpha|^2}(2+e^{i|\alpha|^2})(1+e^{|\alpha|^2}+e^{\frac{1+i}{2}|\alpha|^2})\notag\\
    \chi=&3e^{-2|\alpha|^2}(1+e^{\frac{3-i}{2}|\alpha|^2}[1+e^{i|\alpha|^2}])&
\end{align}
Then, to find an orthonormal basis, we diagonalize $\eta_{qq'}$, obtaining 4 positive-definite eigenvalues:
\begin{equation}
        \lambda_1 =\rho+\chi+6\Re{\tau},
\end{equation}
with degeneracy 1, 
\begin{equation}
        \lambda_2 =\rho-\chi+2\sqrt{3}|\Im{\tau}|,\quad
        \lambda_3 =\rho-\chi-2\sqrt{3}|\Im{\tau}|,
\end{equation}
which each have degeneracy 2, and
\begin{equation}
        \lambda_4 =\rho+\chi-2\Re{\tau},
\end{equation}
with degeneracy 3. From this, we can associate the eigenstates with the logical states of our quoctit:
\begin{equation}
    \ket{\bar{n}}=\sum_{q\in Q}n_q \ket{\phi_q}.
\end{equation}
In this way, $\ket{\bar{0}}$ is associated with $\lambda_1$, $\ket{\bar{1}},\ket{\bar{2}}$ with $\lambda_2$, $\ket{\bar{3}},\ket{\bar{4}}$ with $\lambda_3$, and $\ket{\bar{5}},\ket{\bar{6}},\ket{\bar{7}}$ with $\lambda_4$. 
The $\{n_q\}$ can then be represented as vectors in the ordered basis $\{ \ket{\phi_{+1}},\ket{\phi_{+\bfj}},\ket{\phi_{+\bfi}},\ket{\phi_{+\bfk}},\ket{\phi_{-1}},\ket{\phi_{-\bfj}},\ket{\phi_{-\bfi}},\ket{\phi_{-\bfk}}\}$:
\begin{align}
    \ket{\bar{0}}&=\nu_0^{-1}\times(1,1,1,1,1,1,1,1)\notag\\ 
    \ket{\bar{1}}&=\nu_1^{-1}\times(\kappa_{--}, -\kappa_{+-},1,i,-\kappa_{--},\kappa_{+-},-1,-i)\notag\\
    \ket{\bar{2}}&=\nu_2^{-1}\times(-\kappa_{++},\kappa_{-+},1,-i,\kappa_{++},-\kappa_{-+},-1,i)\notag\\
    \ket{\bar{3}}&=\nu_3^{-1}\times(\kappa_{+-}, -\kappa_{--},1,-i,-\kappa_{+-},\kappa_{--},-1,i)\notag\\
    \ket{\bar{4}}&=\nu_4^{-1}\times(-\kappa_{-+},\kappa_{++},1,i,\kappa_{-+},-\kappa_{++},-1,-i)\notag\\
    \ket{\bar{5}}&=\nu_5^{-1}\times(1,i,-1,-i,1,i,-1,-i)\notag\\
    \ket{\bar{6}}&=\nu_6^{-1}\times(1,-1,1,-1,1,-1,1,-1)\notag\\
    \ket{\bar{7}}&=\nu_7^{-1}\times(1,-i,-1,i,1,-i,-1,i)
\end{align}
with $\kappa_{ab}=\frac{i\pm1}{2}(\sqrt{3}\pm1)$ where $a,b$ indicate the sign.  The normalization of these modes are given by $\nu_i$: 
\begin{align}
     \nu_0 = \bigg(48e^{-\alpha^2}\bigg[2&
                    +8\cos\left(\frac{\alpha^2}{2}\right)\cosh\left(\frac{\alpha^2}{2}\right)
                    \notag\\&+\cos\left(\alpha^2\right)+\cosh\left(\alpha^2\right)\bigg]\bigg)^{\frac{1}{2}}.
\end{align}

For $\nu_i$ with $i=\{1,2,3,4\}$, we define them via
\begin{widetext}
\begin{align}
\nu_{abc} = \left(48e^{-\alpha^2}
                \left[
                    \cos\left(\frac{\alpha^2}{2}\right)
                    -\cosh\left(\frac{\alpha^2}{2}\right)
                \right]
                \left[
                    \left( \pm 1 \pm \sqrt{3} \right)
                    \sin\left(\frac{\alpha^2}{2}\right)
                    -\left(3 \pm \sqrt{3} \right)
                    \sinh\left(\frac{\alpha^2}{2}\right)
                \right]\right)^{\frac{1}{2}}
\end{align} 
\end{widetext}
where $a,b,c$ indicate the sign of each $\pm$. With this:
\begin{equation}
    \nu_1 \equiv \nu_{-+-}\quad \nu_2 \equiv \nu_{+++}\quad \nu_3 \equiv \nu_{+--}\quad \nu_4 \equiv \nu_{--+}
\end{equation}
then for the final 3 $n_i$,
\begin{equation}
\nu_{5},\nu_6,\nu_7 = 4(e^{-\alpha^2}
            \left[3\cosh(\alpha^2)-2-\cos(\alpha^2)\right])^{\frac{1}{2}} 
\end{equation}

\noindent Relationships can be found within the eigenspaces because the orthogonality of the eigenspaces of the Gram matrix is determined by the \textit{subgroups} and \textit{cosets} of the $2T$. $\operatorname{span} \left\{ \ket{\Bar{0}} \right\}$ is spanned by ``the subgroup $Q_8$". For $\operatorname{span} \left\{ \ket{\Bar{1}}, \ket{\Bar{2}}, \ket{\Bar{3}}, \ket{\Bar{4}} \right\}$ is the Hilbert space:
\begin{align}
\operatorname{span}
            \{ &
                \ket{\phi_{+1}}-\ket{\phi_{-1}},
                \ket{\phi_{+\bfi}}-\ket{\phi_{-\bfi}},\notag\\
                &\ket{\phi_{+\bfj}}-\ket{\phi_{-\bfj}},
               \ket{\phi_{+\bfk}}-\ket{\phi_{-\bfk}}
            \},
        \end{align}
with constellations $\left\{ \ket{\phi_q},\ket{\phi_{q'}} \right\}$ composed by the states associated to the quaternions $\{q,q'\}$, representing \textit{any} of the 4 possible \textit{cosets} of $Q_8$ of order 2.
Meanwhile, $\operatorname{span} \left\{ \ket{\Bar{5}}, \ket{\Bar{6}}, \ket{\Bar{7}} \right\}$ is spanned by all the \textit{subgroups} of order 4 of $Q_8$:
\begin{align}
            \operatorname{span}
            \{ 
                & \ket{\phi_{+1}}
                    -\ket{\phi_{+\bfi}}
                    +\ket{\phi_{-1}}
                    -\ket{\phi_{-\bfi}}, \\
                & \ket{\phi_{+1}}
                    -\ket{\phi_{+\bfj}}
                    +\ket{\phi_{-1}}
                    -\ket{\phi_{-\bfj}}, \\
                &  \ket{\phi_{+1}}
                    -\ket{\phi_{+\bfk}}
                    +\ket{\phi_{-1}}
                    -\ket{\phi_{-\bfk}}
            \},
\end{align}

We can similarly ask what the effect of the Gaussian passive transformation $\mathcal{U}$ has on our logical states.  Taking Eq.~(\ref{eq:uzgo}) we see that the effect is to permute the coefficients of the basis states in $\ket{\bar{n}}$.  Therefore it is not simply a $\bar{Z}$.  This is to be expected given the qudit has $Q_8$ rather than $Z_8$ group structure.  Working through, the effect of this transformation on the logical states is
\begin{widetext}
\begin{equation}
\mathcal{U}_8=
\begin{pmatrix}
1 & 0 & 0 & 0 & 0 & 0 & 0 & 0 \\
0 & \frac{1-\sqrt{3}}{4}-\frac{1+\sqrt{3}}{4}i & \frac{(1-\sqrt{3}i)}{2\sqrt{2}} & 0 & 0 & 0 & 0 & 0 \\
0 & -\frac{(1-\sqrt{3}i)}{2\sqrt{2}} & \frac{1+\sqrt{3}}{4}+\frac{1-\sqrt{3}}{4}i & 0 & 0 & 0 & 0 & 0 \\
0 & 0 & 0 & \frac{1-\sqrt{3}}{4}+\frac{1+\sqrt{3}}{4}i  & -\frac{(1-\sqrt{3}i)}{2\sqrt{2}} & 0 & 0 & 0 \\
0 & 0 & 0 & \frac{(1-\sqrt{3}i)}{2\sqrt{2}} & \frac{1+\sqrt{3}}{4}-\frac{1-\sqrt{3}}{4}i & 0 & 0 & 0 \\
0 & 0 & 0 & 0 & 0 & -\frac{i}{2} & \frac{1}{2} + \frac{i}{2} & \frac{1}{2} \\
0 & 0 & 0 & 0 & 0 & \frac{1}{2} + \frac{i}{2} & 0 & \frac{1}{2} - \frac{i}{2} \\
0 & 0 & 0 & 0 & 0 & \frac{1}{2} & \frac{1}{2} - \frac{i}{2} & \frac{i}{2}
\end{pmatrix}.
\end{equation}
\end{widetext}
Comparing this to the $2T$-qutrit, we observe that rather the performing a phase gate, the effect is to perform rotations within the eigenvalue sectors.  Further, one observes that $(\mathcal{U}_8)^3=\mathbb{1}$.  Future work should identify the necessary additional gates to create a universal quoctit gate set.
\section{Robustness to Noise}
\label{sec:numresult}
\noindent A cavity-based quantum processor confronts many potential errors that can impact its performance. These errors arise from various sources: thermal fluctuations in the environment, amplifier noise originating from measurement components, and imperfections in the system's control fields. The interconnected nature of components introduces cross-correlation noise, adding another layer of complexity to error considerations. Here, we focus on two common errors: pure-loss and dephasing.

Pure-loss errors manifest in various ways, often attributed to factors such as energy dissipation in Josephson junctions, microscopic defects or impurities in fabrication materials, and intrinsic dielectric or conductor losses within superconducting resonators. On the other hand, dephasing errors revolve around the loss of phase information within the quantum system with diverse origins including magnetic flux noise resulting from ambient magnetic fields, unpredictable charge fluctuations, and non-linearities within the Josephson junctions.

Achieving quantum error correction without the need for active measurement and feedback operations can been realized through the implementation of the basic cat-code. However, this code alone does not offer adequate protection against photon loss, as demonstrated in \cite{leghtas2015confining}. In this context, we conduct a comparative analysis of our $2T$-quoctit to other existing codes. Our aim is to understand the benefits and tradeoffs from this dual-mode, nonabelian qudit approach. To evaluate this robustness, we consider the pure-loss channel and the dephasing channel which, when considering its action only on one mode, are described by the following discrete set of Kraus operators \cite{leviant2022quantum}:
\begin{equation}
\left\{
        \hat{K}^p_l = \sqrt{\frac{\gamma^l}{l!}} \;(1-\gamma)^{\hat{n}/2} \hat{a}^l
    \;\;\left|\right.\;\; {l\in\mathbb{N}\cup \{0\}} \right\}
\label{eq:loss}\end{equation}
\begin{equation}
\left\{
        \hat{K}^d_l =
        \sqrt{\frac{\delta^l}{l!}} \exp\left(-\frac{\delta}{2} \hat{n}^2\right)\hat{n}^l
    \;\;\left|\right.\;\; {l\in\mathbb{N}\cup \{0\}}  \right\}
\label{eq:dephasing}\end{equation}
where $\gamma\equiv 1-e^{-\chi}$, $\chi$ is the damping parameter, and $\delta >0$ is the dephasing strength. $\hat{K}^p_l$ for $l>0$ induce the pure-loss errors, while the operators $\hat{K}^p_0$ and $\hat{K}^d_l$ for $l\geq 0$ induce the dephasing errors.
With these, the noise channels are defined as:
\begin{equation}
    \mathcal{N}\,:\,\rho \mapsto \sum_{i=0}^{\infty}\hat{K}_i\rho \hat{K}^\dagger_i,\;\;\;\;\;\;\hat{K}\in \{\hat{K}^p,\hat{K}^d\}
\end{equation}
for some density matrix $\rho$. In this work, we assume the two modes have identical and independent errors, such that we can describe the 2-mode error channel as $\mathcal{N}\otimes\mathcal{N}$.

To evaluate the performance of our $2T$-quoctit against different channels $\mathcal{C}$, we compute the \textit{entanglement fidelity} following~\cite{Denys:2022iyj}.  This fidelity is used to determine how ``close" a logical entangled state is after applying a channel $\mathcal{C}$ to the second  qudit.

\vspace{2mm}
\noindent 
To do this, we define a maximally-entangled state of two logical qudits $\ket{\Phi_d}\coloneqq \frac{1}{\sqrt{d}}\sum_{k=0}^{d-1}\ket{k}\!\ket{k}\in \mathbb{C}^d\otimes\mathbb{C}^d$, (where  $d=3$ ($d=8$) for the qutrit (quoctit) system). The entanglement fidelity is defined by:
\begin{equation}
    F(\mathcal{C})\coloneqq
    \bra{\Phi_d}
    [\operatorname{Id}\otimes\,
    \mathcal{C}]\left(
        \ket{\Phi_d}\!\bra{\Phi_d}
    \right)
    \ket{\Phi_d},
\end{equation}
where the channel $\mathcal{C}=\mathcal{R}\circ\mathcal{N}\circ\mathcal{E}$ is a composition of three successive maps: the \textit{encoding} map $\mathcal{E}$, which takes the computational basis  $\ket{k}$ to the codespace basis $\ket{\bar{k}}$.  For the $2T$-qudits this basis is the $2T$ constellation (thus, its Kraus operators will be $24\times d$ matrices);
the noise channel map $\mathcal{N}$; and the recovery map $\mathcal{R}$ which we use to reconstruct as much of the original logical state as possible.

For initializing $\mathcal{E}$ we take either a $2T$-qudit, $dn-PSK$, or a $d$-dimensonal random encoding. Denoting the Kraus operators of the encoding map, noise channel and recovery map as $\set{E_k}, \set{N_k}$ and $\set{R_k}$ respectively, numerical optimization is performed iteratively for $\mathcal{R}$ and $\mathcal{E}$ in that order, in order to maximize the fidelity under a fixed $\mathcal{N}$. This leads to an approximation of the true $\max_{E_k,R_k}F(\mathcal{C})$ with reasonable run time. Moreover, the task is made more tractable because $\mathcal{E},\mathcal{R}$ are semidefinite positive maps, hence the optimization procedures with respect to them are semidefinite programs, which are efficiently solvable via a modified Python code~\cite{Denys:2022iyj} which uses the Splitting Conic Solver package~\cite{odonoghue_conic_2016}.

\subsection{Pure-Loss Tolerance Comparison}

\noindent The pure-loss channel action on coherent states is described by a \textit{damping} of the type $\ket{\alpha}\mapsto\ket{(1-\gamma)^{1/2}\alpha}$ (with $\gamma = 1-e^{-\chi}$).
Since our constellation  comprises of  24 (2-mode) states, we need 24 Kraus operators to describe the action of the channel.

\begin{figure}
    \centering
    \includegraphics[width=\linewidth]{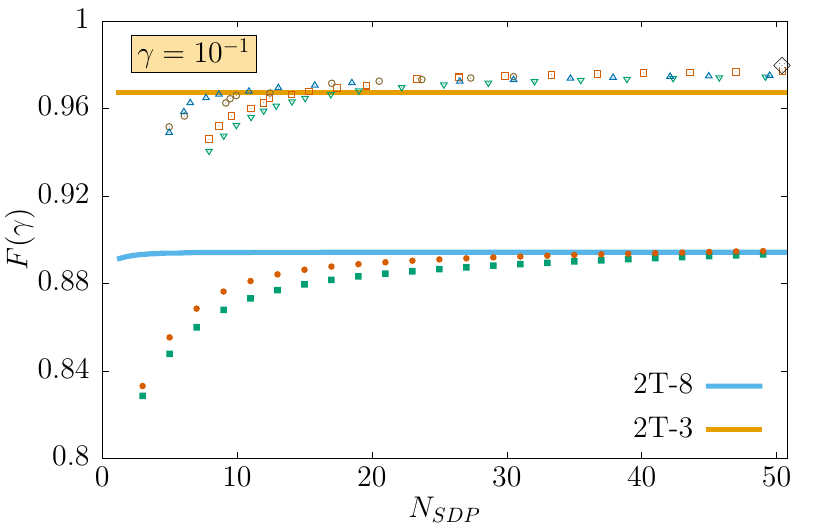}\\
    \includegraphics[width=\linewidth]{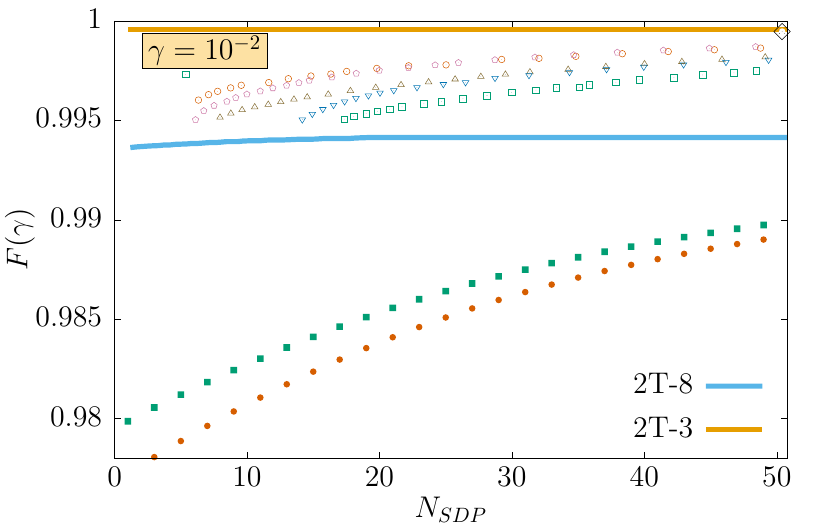}
    \caption{$F(\gamma)$ as function of the number of optimization steps for two values of $\gamma$.
    The lines are for the $2T$-qudit encodings, while the solid (empty) points are for $d=8$ ($d=3$) random encodings.
    $\alpha$ is chosen to be near the $2T$ optimal values: $\alpha=2.0$ for the quoctit encodings, while $\alpha=1.5$ for the qutrit encodings from~\cite{Denys:2022iyj}.}
    \label{fig:sdp}
\end{figure}

In Fig.~\ref{fig:sdp} we show the results of the fidelities as a function of semi-definite programming optimization steps $N_{SDP}$ when starting from different encodings: the $2T$-quoctit, the $2T$-qutrit and randomly encoded qudits for a fixed $\alpha$ for each one of the plots.
We can see that similar to the previous results for $2T$-qutrit, encoding in a $2T$-quoctit has nearly optimal $F(\gamma)$ and require far smaller $N_{SDP}$ compared to the random encodings. This suggests that these are close to the optimal encodings inside the $2T$ constellation. While the $2T$-quoctit outperforms the random 8-level encodings, it is lower than the 3-level random encodings and the $2T$-qutrit.

This is expected as the $2T$-quoctit contains more logical information inside the $2T$ constellation and thus the remaining error correcting space is smaller. Therefore it is important to look at exactly \emph{how much} the fidelity has been reduced.  We will return to this quantitative question as a function of $\gamma$ after optimizing $\alpha$.

\begin{figure}
    \centering
    \includegraphics[width=\linewidth]{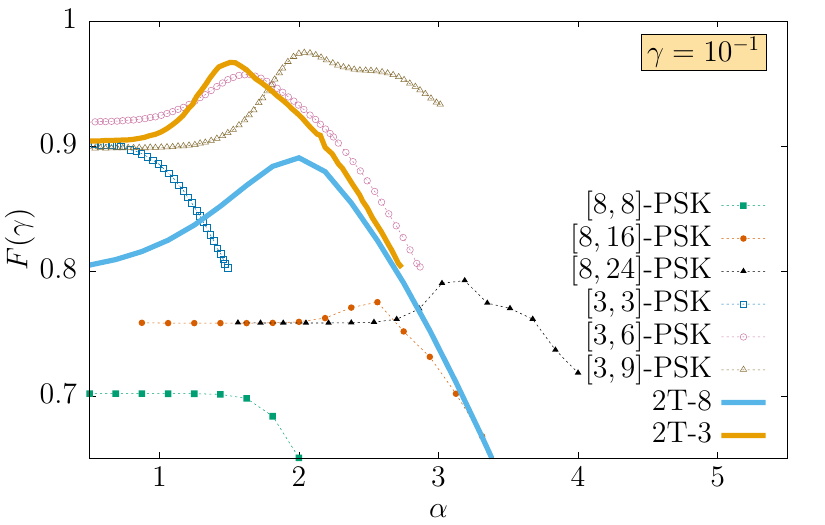}\\
    
    \includegraphics[width=\linewidth]{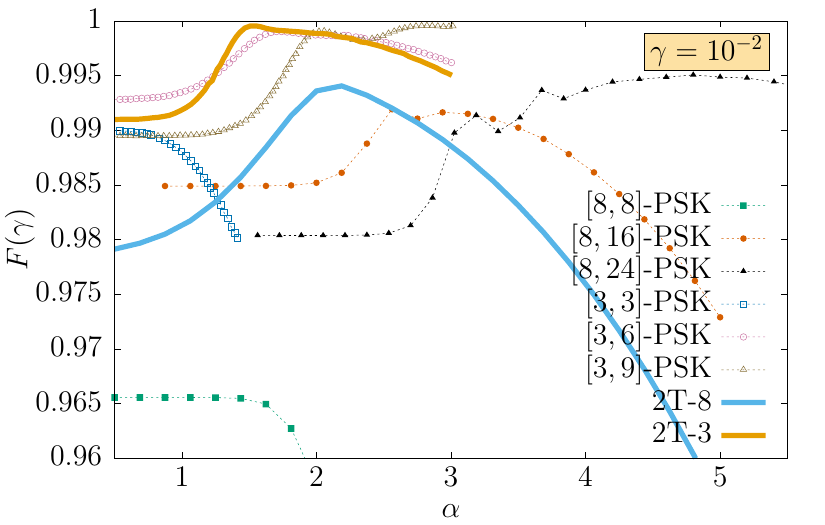}
    \caption{$F(\gamma)$ as a function of $\alpha$. Data for qutrit encodings from~\cite{Denys:2022iyj}}
    \label{fig:alpha}
\end{figure}

\begin{figure}[h]
    \centering
    \includegraphics[width=1\linewidth]{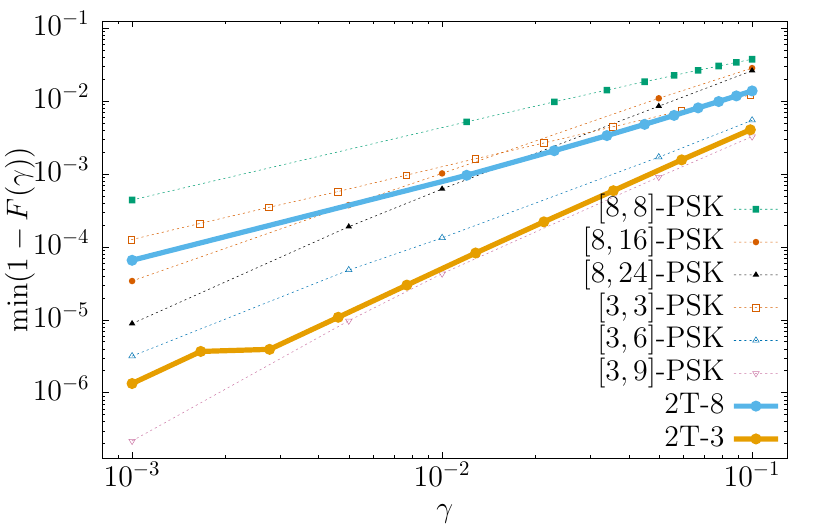}
    \caption{Infidelity as a function of $\gamma$ for optimal $\alpha$.}
    \label{fig:smallest_infid}
\end{figure}

\noindent We repeated these calculations for a range of $\alpha$ and $\gamma$. In Fig.~\ref{fig:alpha} we plot the entanglement fidelity for different qudit encodings as a function of the coherence parameter $\alpha$ for tthe values  $\gamma=10^{-1},10^{-2}$. The optimal $\alpha$ for the $2T$-quoctit is found to be $\approx 2$ and appears relatively insensitive to $\gamma$.  This is larger then the previously determined value of $\alpha\approx1.5$ found for the $2T$-qutrit~\cite{Denys:2022iyj}. In contrast to the $2T$-qudits, the optimal $\alpha$ for the [d,n]$-$PSK qudits grows rapidly with $\gamma$.  For the $\gamma$ studied here, optimal $\alpha$ grew as large as 12. For the optimal $\alpha$, we present in Fig.~\ref{fig:smallest_infid} the minimum entanglement infidelity $\text{min}(1-F(\gamma))$ as a function of $\gamma$.

Fitting to the curves, we find for the $2T$-qudits that their fidelities as a function of $\gamma$ are
\begin{align}
    \mathcal{F}_{2T-8}(\gamma)&=1-1.459\gamma^{1.163}\\
    \mathcal{F}_{2T-3}(\gamma)&=1-1.662\gamma^{1.776}
\end{align}

From the results in Fig.~\ref{fig:smallest_infid}, we see that for sufficiently small $\gamma$ and large $n$, $[d,n]$-$PSK$ have lower infidelity than $2T$-qudits when $\alpha$ is unrestricted. In particular $[8,16]$-$PSK$, $[8,24]$-$PSK$ and $[3,9]$-$PSK$ all achieve lower infidelity around $\gamma=10^{-2}$ compared to their respective $2T$-qudit. In order to achieve these low infidelities, the optimal $\alpha$ for $[d,n]$-$PSK$ grows inversely with $\gamma$, and for $\gamma=10^{-3}$ can exceed $\alpha>12$.

In SRF-based systems, while the constraints are not as strict as for transmons, it is possible for the cavity drive to not be simultaneously sufficiently weakly-coupled, able to deliver sufficient power for large $\alpha$, and fast without deleterious heating or crosstalk.~\cite{Roy:2024uro}. Recently SRF-based systems have demonstrated coherent states with $\alpha=\sqrt{20}\approx 4.47$~\cite{Kim:2025ywx}, which given the average number of photons is $n=|\alpha|^2$ corresponds to 20 photons.  

Upper bounds on $\alpha$ can be obtained by considering the critical photon number $|\alpha_\kappa|^2=n_\kappa=\frac{\kappa\Delta}{\chi^2}$ where $\kappa$ is the photon decay rate, $\Delta=\omega_a-\omega_r$ is the cavity-qubit detuning, and $\chi$ is the Stark shift per photon. When $n\geq n_\kappa$, an instability occurs where the dispersive coupling is lost and the qubit becomes strongly coupled to the cavity states~\cite{schuster2007circuit}.  For the SRF-based systems in \cite{Kim:2025ywx}, $\alpha_\kappa\geq 100$, although high-fidelity control of such large coherent states remains to be demonstrated, and other limitations may further upper bound viable $\alpha$.

\begin{figure}[h]
    \centering
    \includegraphics[width=1\linewidth]{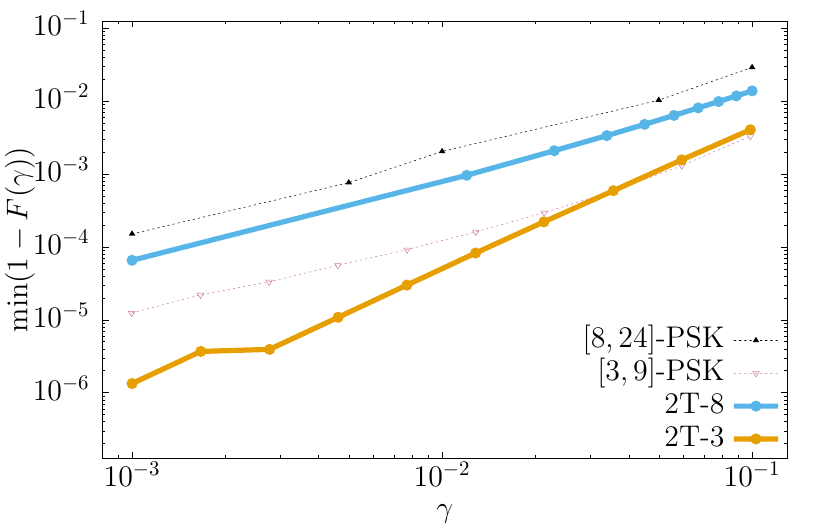}
    \caption{Infidelity as a function of $\gamma$ for optimal $\alpha$ when restricting $\alpha\leq 3$.}
    \label{fig:smallest_infid_res}
\end{figure}

With these experimental prospects in mind, it is clear that depending on the range of $\alpha$ that allows for high-fidelity control, the relative efficiency of $[d,n]$-$PSK$ vs $2T$-qudits can vary.  If we restrict to $\alpha\leq 3$, the $2T$-qudits outperform all their respective $[d,n]$-$PSK$ qudits (See Fig.~\ref{fig:smallest_infid_res}).  Starting at $\alpha=4$, the $[8,24]$-$PSK$ quoctit outperforms the $2T$-quoctit for $\gamma<10^{-2}$. 

Given the difference in qudit dimension, it is unsurprising that the infidelity for the $2T$-quoctit, $1-\mathcal{F}_8$ is higher than the $2T$-qutrit, $1-\mathcal{F}_3$.  One can ask to what extent it is worse relative of fixed quantum information.  If the decrease in fidelity was entirely an effect of reduced redundancy, one would naively expect the infidelity to be \begin{equation}
    \mathcal{I}_R=\frac{1-\mathcal{F}_8}{1-\mathcal{F}_3^{\log_3(8)}}\approx 1.
\end{equation} Fig.~\ref{fig:smallest_infid_res} shows that the relative infidelity increases as $\gamma\rightarrow0$, and while accounting for the factor $\log_3(8)$ does reduce $\mathcal{I}_R$, one still finds $\mathcal{I}_R\gg 1$ as the noise is reduced.  While this result reduces the usefulness of $2T$-quoctit over $2T$-qutrits, it is not necessarily fatal.  This reduced fidelity per unit of quantum information could be overcome by reductions in circuit depth: in particular, from fewer entangling gates.

\subsection{Dephasing Tolerance Comparison}

From \Cref{eq:dephasing}, one observes that the dephasing channel mixes coherent states of the constellation. Moreover, it does not even map coherent states to pure states. Thus, to model the dephasing channel there is  need to simulate the system in a larger Hilbert space.  One choice is to represent things in  the Fock representation, with the effect of the noise channel given by
\begin{equation}
\mathcal{N}[\gamma_d](\hat{\rho})=\sum_{m,n=0}^\infty e^{-\frac{1}{2}\gamma_d(m-n)^2}\braket{m|\hat{\rho}|n}\ket{m}\bra{n}.
\label{eq:dephasing_fock}\end{equation}
The noise channel consists of infinite number of Kraus operators, and thus we must impose a numerical truncation unlike the pure-loss channel. Further, we will assume that the noise channels are independent and equal in each mode, allowing us to consider the two-mode generalization of Eq.~(\ref{eq:dephasing_fock}) for our qudits.

\begin{figure}
    \centering
    \includegraphics[width=\linewidth]{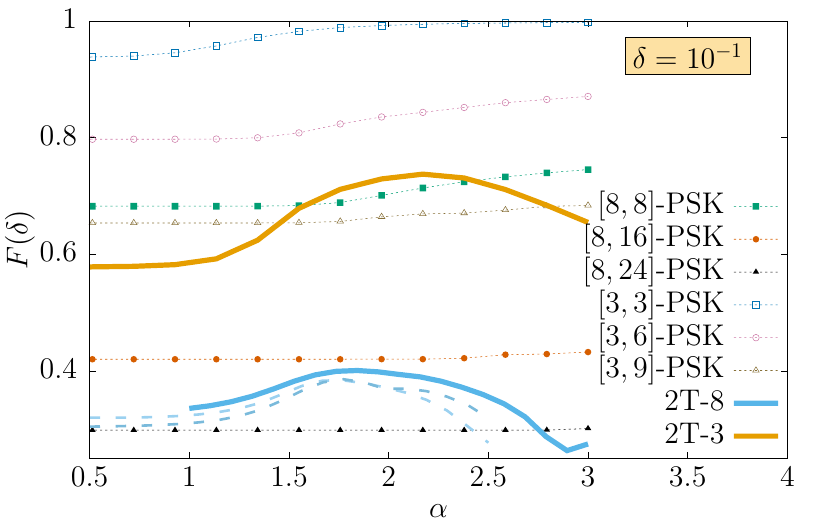}\\
    \includegraphics[width=\linewidth]{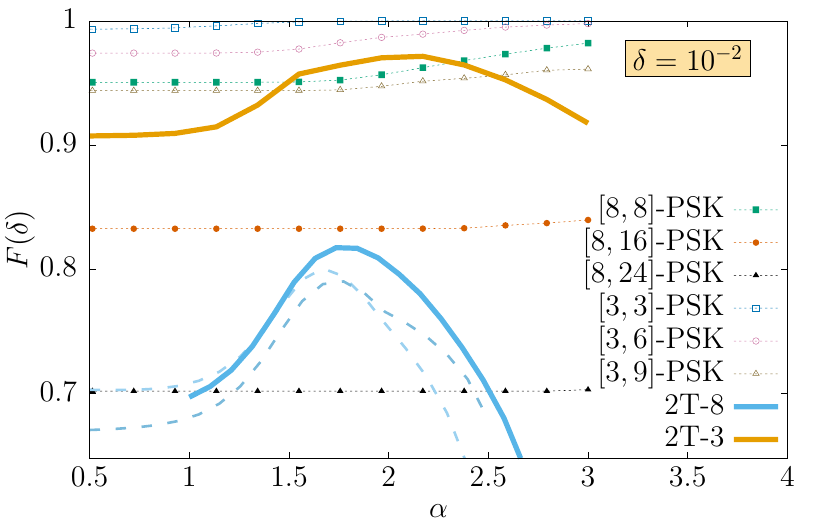}
\caption{$F(\delta)$ as a function of $\alpha$ for the dephasing channel. The different $2T$-quoctit lines indicate different truncations.}\label{fig:alpha_trunc}
\label{fig:dephasing}\end{figure}

In our simulations, we truncate in total number of photons $N=2n$ photons in the two modes i.e. at most $n$ photons in each mode. 
For the $[d,n]$-$PSK$ qudits, $N=40$ are possible because the encoding can be implemented efficiently.  In contrast, the $2T$-qudit simulations do not admit such encodings and thus end up being memory-intensive, restricting us to $N\leq 16$. For all $[d,n]$-$PSK$ and the $2T$-qutrit, this leads to negligible truncation errors.  For the $2T$-quoctit, we find moderate dependence between $N=12,14,16$ with increasing entanglement fidelity $F(\delta)$ with an increasing $N$.  From this, we understand our numerical results for $2T$-quoctit to be lower bounds on the true entanglement fidelity -- which can be investigated experimentally. We plot the results of simulating this channel in Fig.~\ref{fig:dephasing}. For the $2T$-quoctit, we present the $N=12,14,16$ results with a solid line for the $N=16$ case.  For all other qudits, we show only the largest truncation results where convergence is observed.

We observe that for all $\alpha$ $[d,d]$-$PSK$ and $[d,2d]$-$PSK$ qudits perform better than $2T$-qudits, while the $[d,3d]$-$PSK$ generically perform worse.  For the $2T$-quoctit, at optimal $\alpha$ appears to perform within 3\% of the $[8,16]$-$PSK$ qudit.  Given the lack of convergence observed for the $2T$-quoctit, it is possible that it outperforms the $[8,16]$-$PSK$ qudit and should be a focus of future investigation either numerically or experimentally. Further, different qualitative behavior occurs for the $[d,n]$-$PSK$ and $2T$-qudits.  The PSK qudits have continued improvement in fidelity as $\alpha$ is increased, while the $2T$-qudits are strongly peaked near $\alpha=2$.

These preliminary results for the $2T$-quoctit are promising, especially given that we have assumed that both modes are suffering from independent phase noise.  As pointed out in~\cite{Denys:2022iyj}, with actual devices this is potentially too pessimistic and one might expect correlations between the noise of the two modes, and therefore higher performance would be achieved.

\subsection{Combined Fidelities}
\noindent Real bosonic platforms will have to address both noise channels to some degree.  Given that the $\alpha$ dependence of the fidelities in each channel are different, the optimal $\alpha$ for a given platform should weigh the relative error rates to obtain the best overall fidelity.

\begin{figure}
     \centering
    \includegraphics[width=\linewidth]{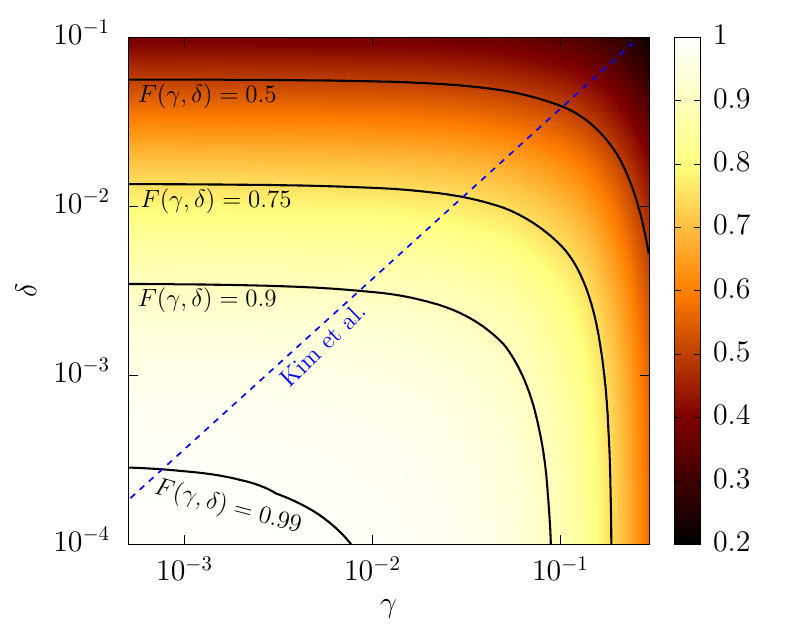}
    \caption{Maximum fidelity under combined noise channels $F(\gamma,\delta)$ for a $2T$-quoctit as a function of $\gamma,\delta$.\label{fig:joint_fid}  The dashed blue line indicates the average experimental $T_1$ and $T_\phi$ found in~\cite{Kim:2025ywx} for a two-mode SRF cavity device. The exact location on the line obtainable for a $2T$-quoctit depends on the error correction cycle time as discussed in the text.}
\end{figure}

Here, we take a simple model where the two errors are uncorrelated and occur sufficiently rarely that the overall fidelity can be approximated as a product of the two fidelities, and then determine the maximum value:

\begin{equation}
    F(\gamma,\delta)=\max_{\alpha}[F(\gamma,\alpha)F(\delta,\alpha)],
\end{equation}
which we present in Fig.~\ref{fig:joint_fid} for reference in benchmarking device performance. An important observation from the combined scenario is that the fidelity is nearly an order of magnitude more sensitive to dephasing rate compared to depolarizing rate.  Further, while $\mathcal{F}(\gamma,\delta)=0.99$ may seem small, one should remember that the effective qubit fidelity is enhanced to $\mathcal{F}(\gamma,\delta)^{1/\log_2(8)}=0.997$.

The two channels have different $\alpha$ dependence.  $F(\gamma)$ is strongly peaked around $\alpha=2$, while $F(\delta)$ peaks around $1.8$ albeit with weaker $\alpha$ dependence. While this introduces a competition between the two channels in practice, it is milder than the $PSK$ qudits where the optimal values diverge with decreasing $\gamma,\delta$. For the $2T$-quoctit, we find that under the combine noise model the optimal $\alpha\approx 1.85$ varying by only 8\% over the range of $\gamma,\delta$ considered.  This implies that while our noise model is simplistic, it should be robust to more realistic error models.

With these results, it is possible to make comparisons to exisiting and potential bosonic platforms. Kim et al. presented an experimental demonstration of a two-mode cavity with depolarizing times $T^A_1=20.6$ ms \& $T^B_1=15.6$ ms, and pure dephasing times $T^A_\phi=43.2$ ms \& $T^B_\phi=59.8$ ms where $A,B$ denote each of the two modes. These properties can be mapped to $\gamma=1-e^{-T/T_1}$ and $\delta=T/T_\phi$ respectively for a fixed time $T$ to perform the error detection and recovery step. This parameterized line is included in Fig.~\ref{fig:joint_fid} and can be used to estimate the maximum time for an error correction cycle to obtain a fixed $\mathcal{F}(\gamma,\delta)$.  For example, with these $T_1,T_\phi$ and $\mathcal{F}(\gamma,\delta)=0.9$ the error correction cycle must last no longer than $4\times 10^{-3}\times15.6$ ms = 60 $\mu s$.

\section{Conclusions}
\label{sec:con}
In this work, we developed a $Q_8$ group-structured quoctit inside a $2T$ constellation of coherent states in multimode bosonic systems.
Using a modified version of previously-developed code used for studying $2T$-qutrits~\cite{Denys:2022iyj}, we investigated the error tolerance of the $2T$-quoctit encoding against pure-loss and dephasing channels. We found that the $2T$-quoctit can outperform PSK qudits when restricting the total power in the system.   While the $2T$-quoctit does not have comparable fidelity to the $2T$-qutrit at fixed quantum information, there is reason for hope given potential reductions in circuit depth or other sources of qudit advantage.  Additional future work should focus on more realisitic noise modeling and experimental demonstration of the $2T$-qudits.  

Having demonstrated the ability to construct viable non-abelian qudits in multimode cavities, an obvious extension would be to a \textit{quicosotetrit} (a qudit with $d=24$) with $2T$ group structure built out of a larger group such at $2O$ or $2I$.  Having such qudits would enable the exploration of fault-tolerant simulations of lattice gauge theories on specialized platforms. Additional research would focus on establishing a universal gate set for these qudits and the other necessary components to a fully fault-tolerant quantum computer.  Such information would then lead to concrete resource estimations for the relative merits of different qudits to applications e.g. particle physics.

\begin{acknowledgements}
The authors would like to thank Tanay Roy and Yao Lu for a number of useful discussions and comments in the process of completing this work. This material is based on work supported by the U.S. Department of Energy, Office of Science, National Quantum Information Science Research Centers, Superconducting Quantum Materials and Systems Center (SQMS) under contract number DE-AC02-07CH11359. This work was produced by Fermi Forward Discovery Group, LLC under Contract No. 89243024CSC000002 with the U.S. Department of Energy, Office of Science, Office of High Energy Physics. 
\end{acknowledgements}

\bibliography{wise}

\end{document}